\documentclass[12pt]{iopart}
\usepackage{citesort} 
\usepackage{siunitx}
\usepackage{floatflt}
\usepackage{graphicx}
\usepackage{pdfpages}
\usepackage[utf8]{inputenc} 
\usepackage[T1]{fontenc} 
\usepackage{floatflt}

\begin{document}

\title[A morphology study on the epitaxial growth of graphene and its buffer layer]{Tailoring the SiC surface - a morphology study on the epitaxial growth of graphene and its buffer layer}

\author{Mattias Kruskopf$^1$(*), Klaus Pierz$^1$, Davood Momeni Pakdehi$^1$, Stefan Wundrack$^1$, Rainer Stosch$^1$, Andrey Bakin$^{2,3}$ and Hans W. Schumacher$^1$}

\address{$^1$Physikalisch-Technische Bundesanstalt, Bundesallee 100, 38116 Braunschweig, Germany

$^2$Institute of Semiconductor Technology of Technische Universität Braunschweig, Hans-Sommer-Straße 66, 38106 Braunschweig, Germany 

$^3$Laboratory for Emerging Nanometrology (LENA), TU Braunschweig, Germany
}
\ead{Mattias.Kruskopf@ptb.de, Klaus.Pierz@ptb.de}
\vspace{10pt}
\begin{indented}
\item[]July 2017
\item[]\textbf{Keywords:} epitaxial graphene, buffer layer growth, polymer-assisted, giant step bunching, hydrogen etching
\end{indented}

\begin{abstract}
We investigate the growth of the graphene buffer layer and the involved step bunching behavior of the silicon carbide substrate surface using atomic force microscopy. The formation of local buffer layer domains are identified to be the origin of undesirably high step edges in excellent agreement with the predictions of a general model of step dynamics. The applied polymer-assisted sublimation growth method demonstrates that the key principle to suppress this behavior is the uniform nucleation of the buffer layer. In this way, the silicon carbide surface is stabilized such that ultra-flat surfaces can be conserved during graphene growth on a large variety of silicon carbide substrate surfaces. The analysis of the experimental results describes different growth modes which extend the current understanding of epitaxial graphene growth by emphasizing the importance of buffer layer nucleation and critical mass transport processes.
\end{abstract}
%
%
%
%
%

\section*{Introduction}

Clean crystal surfaces at high temperatures reveal characteristics that may be described by “annealing”, “etching” and “growth” \cite{Vlachos1993,Schwoebel1966,Schwoebel1968,Weeks2007}. In the case of the two-component crystal SiC, these mechanisms may coincide since here the partial pressures of silicon and carbon-containing vapor species need to be considered separately to describe the equilibrium conditions at the surface. Especially below temperatures of \SI{2000}{\degreeCelsius} the low vapor pressure of carbon-containing species compared to that of silicon species is the reason why, for example, thermal etching of the terrace edges may be accompanied by the growth of carbon domains \cite{Raback1999,Tromp2009,Ohta2010}. In the case of annealing, thermally activated species reorganize on the surface without ultimately leaving the surface and enable the formation of an energetically preferred configuration \cite{Schwoebel1968}. In the case of clean SiC (0001) surfaces at temperatures of about \SI{1400}{\degreeCelsius} in an argon atmosphere, this restructuring process leads to the formation of new surface steps but also to the formation of carbon domains due to preferred silicon desorption \cite{Kruskopf2015,Ohta2010}. The first carbon layer is the so-called buffer layer that saturates about $1/3$ of the dangling silicon bonds of the underlying substrate surface forming the $(6\sqrt{3} \times 6\sqrt{3})$R30 reconstruction \cite{Starke2009,Riedl2010b}. Depending on the process parameters and substrate properties, these morphological changes are often accompanied by the formation of terrace structures with heights of several nanometers, so-called giant steps. Especially substrates with a larger miscut angle  ($\geq \SI{0.2}{\degree}$) or those that were initially hydrogen-etched are known to support giant step bunching \cite{Oliveira2011,Virojanadara2009}. Graphene formation across giant steps is typically described by step flow growth leading to the formation of monolayer domains on the terraces but also to continuous multilayer domains along the edges \cite{Ohta2010}. Most of the published experiments on epitaxial graphene growth, solely focus on the development of the graphene layer presuming a SiC surface that is already reconstructed by a buffer layer to explain their proposed models \cite{Borovikov2009,Ohta2010,Yazdi2013,Ming2011}. However, this cannot adequately describe the dependency of giant step formation on process parameters such as the pressure of the ambient inert gas, heating rate, temperature and the properties of the starting surface \cite{Bao2016,Virojanadara2010}. Some recent studies show that substrates may be processed such that low steps (e.g. \SI{0.75}{nm}) are conserved by employing substrates with a small miscut angle ($\leq \SI{0.1}{\degree}$), by suitable annealing sequences or by polymer deposition for improved buffer layer growth \cite{Virojanadara2009,Virojanadara2010,Kruskopf2015,Kruskopf2016}.

\begin{figure}
		\centering
    \includegraphics[width=1\textwidth]{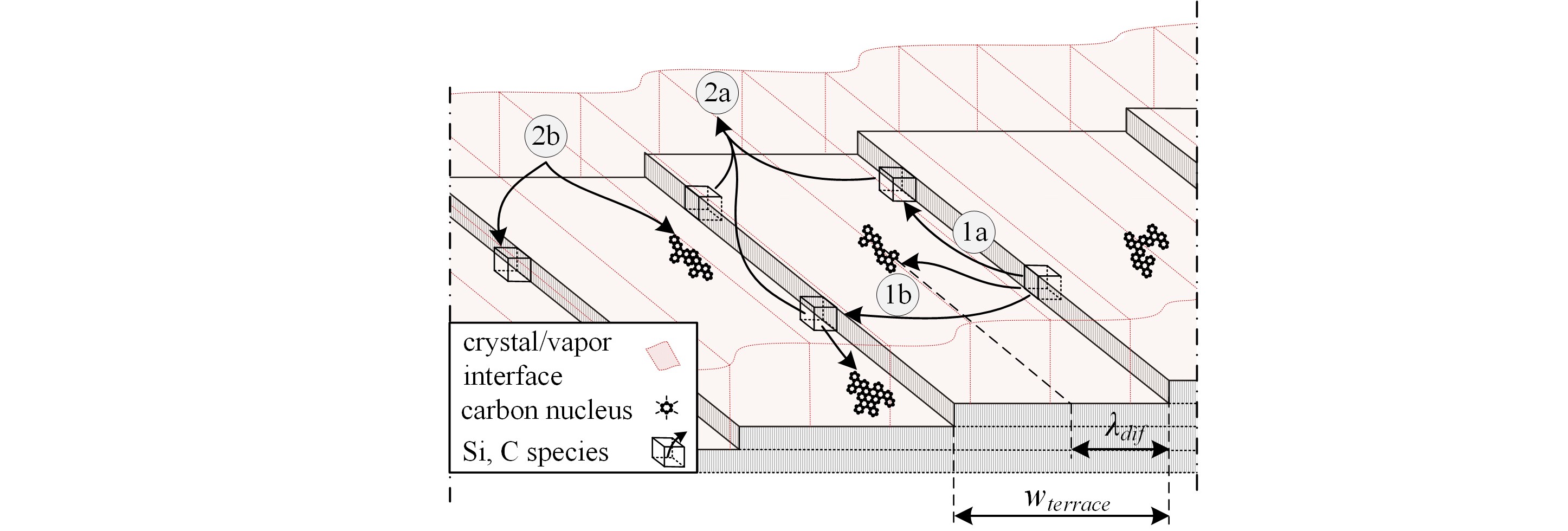}
	  \caption[Mass transport mechanisms during high-temperature annealing]{\textbf{Mass transport mechanisms during high-temperature annealing.} Silicon and carbon-containing species are released once the SiC substrate decomposes. The vapor and diffusing species at the crystal/vapor interface determine the equilibrium state at the surface. Dominating silicon sublimation leads to carbon supersaturation on the surface which favors the formation of buffer layer nuclei. Small stable buffer layer domains as well as terrace edges act as preferred crystallization sites and reduce the effective diffusion length $\lambda_{dif}$ on the terraces. Material between step edges and crystallization sites is exchanged by local transport mechanisms comprising step edge and terrace diffusion (1a, 1b) as well as by global transport involving desorption and adsorption processes (2a, 2b) via the crystal/vapor interface (red shade layer)}
		\label{fig:MassTransportMechanisms}
\end{figure}

While the driving force of faceting of 6H-SiC is known to increase as a function of the miscut angle \cite{Nakagawa2003,Nakajima2005}, understanding the reasons for step bunching during epitaxial graphene growth requires further analysis of the role of carbon layers and the involved mass transport mechanisms. The present work focuses on the morphological changes that occur during the restructuring process of the surface to identify critical aspects which determine the quality of the final graphene sample. Figure \ref{fig:MassTransportMechanisms} shows typical mass transport mechanisms at the interface of the SiC crystal surface during the initial stage of surface restructuring and buffer layer formation. Following the descriptions in literature, the equilibrium conditions at the surface are typically described by local reservoirs and sinks leading to growth and etching \cite{Dhanaraj2010,Weeks1997,Jeong1998,Lui1999}. Here, step edges, as well as the vapor/crystal interface, are assigned as silicon and carbon reservoirs while nuclei and the ambient atmosphere may represent sinks. The basic idea of the model is to differentiate between local mass transport processes by surface diffusion (1a, 1b) and so-called global mass transport processes via desorption and adsorption of silicon and carbon species through the vapor phase (2a, 2b). For epitaxial graphene growth, these mechanisms are often manipulated by affecting the mean free path length and partial pressures at the crystal/vapor interface using an inert argon gas atmosphere \cite{Emtsev2009,Virojanadara2010} or by confinement control \cite{Yang2017,DeHeer2011}. The red shade layer in Figure \ref{fig:MassTransportMechanisms} represents the Knudsen layer which is the highly non-equilibrium interface region in which direct interactions between gas molecules and the surface dominate \cite{Ytrehus1996,Gusarov2002}. In the case of \SI{1}{bar} argon pressure, the thickness of the Knudsen layer is estimated to be a few hundred nanometers \cite{Gusarov2002}. The partial pressures of mostly Si and small quantities of Si$_2$C and SiC$_2$ species forming above the SiC surface at temperatures below \SI{2000}{\degreeCelsius} are known to significantly shift the growth temperature of the buffer layer and graphene towards higher values \cite{Tromp2009,Raback1999}. The morphology study presented in this work identifies decisive mass transport and nucleation mechanisms that determine the step bunching behavior as well as the uniformity of the epitaxially grown carbon layers. 

\begin{figure}
		\centering
    \includegraphics[width=1\textwidth]{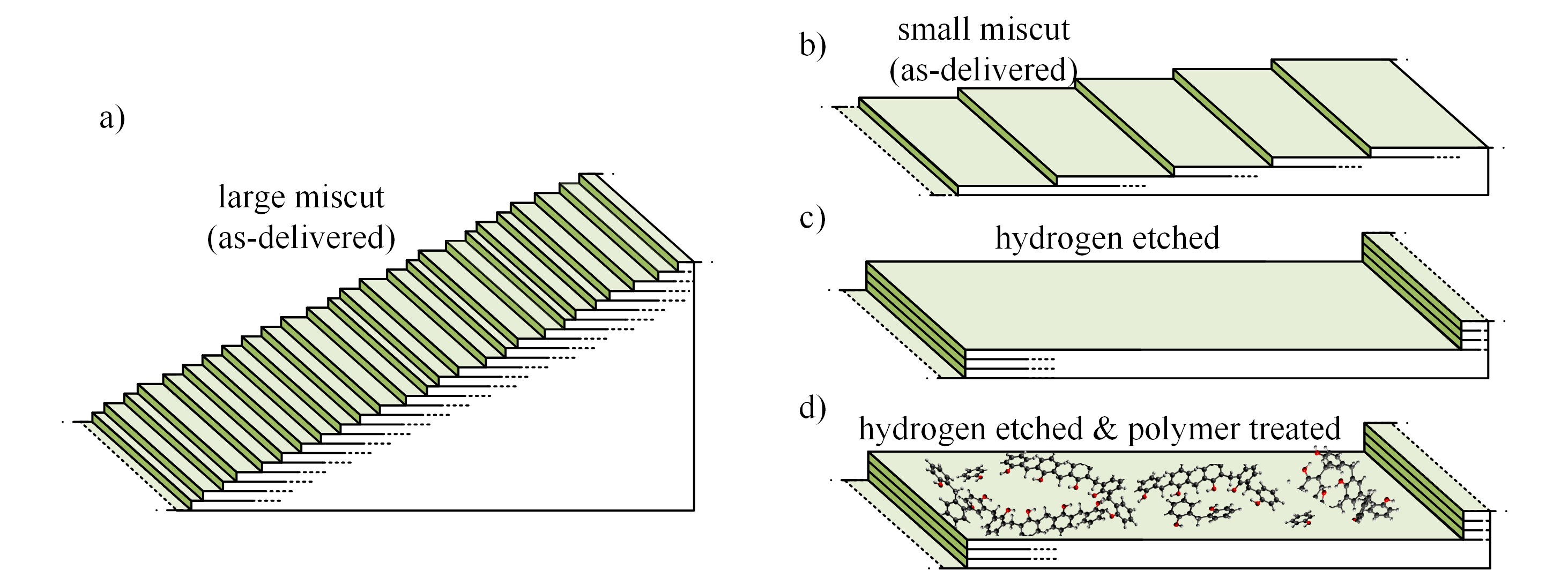}
	  \caption[SiC starting surfaces for buffer layer and graphene growth]{\textbf{SiC starting surfaces for buffer layer and graphene growth.} "As-delivered" substrates with \textbf{(a)} a large miscut angle ($\approx \SI{0.37}{\degree}$) and \textbf{(b)} a small-miscut angle ($\approx \SI{0.05}{\degree}$) provide steps of single SiC bilayers with heights of \SI{0.25}{nm}. \textbf{(c)} Hydrogen etching of small-miscut substrates leads to \SI{0.75}{nm} high terraces corresponding to bunches of three SiC bilayers. \textbf{(d)} Ultimately, hydrogen-etched surfaces are treated with a polymer adsorbate which acts as an additional carbon source for polymer-assisted sublimation growth (PASG) of the buffer layer and epitaxial graphene.}
		\label{fig:StartingSurfaces}
	\end{figure}

The first two sets of experiments (Figure \ref{fig:StartingSurfaces}(a-b)) focus on the influence of the substrate miscut angles. Both surfaces are “as-delivered” with terrace structures of single SiC bilayers and corresponding step heights of \SI{0.25}{nm}. The third set of experiments applies a “hydrogen-etched” surface (Figure \ref{fig:StartingSurfaces}(c)) with predefined steps and terraces having regular heights of \SI{0.75}{nm} which is a stable configuration of the SiC surface for the applied process parameters. In the last set of experiments (Figure \ref{fig:StartingSurfaces}(d)) hydrogen-etched substrates are treated with a polymer to investigate the impact of an additional carbon source on the growth behavior.

\section*{Sample preparation}

The substrates were cut from semi-insulating epi-ready 6H-SiC ($0001$) wafers. The two wafers had different miscut angles of \SI{0.05}{\degree} (small-miscut) and \SI{0.37}{\degree} (large-miscut) with respect to the (0001) crystal plane. The substrates used in the experiments presented in Figure \ref{fig:BufferLayerFormationOnHydrogenEtchedSiC}-\ref{fig:PASGGrapheneOnHydrogenEtchedSiC} were initially prepared by hydrogen etching at temperatures of \SI{1400}{\degreeCelsius} and \SI{1200}{\degreeCelsius}. Etching of the Si-face at \SI{1200}{\degreeCelsius} leads to step heights of \SI{0.25}{nm} and \SI{0.5}{nm}. This is only slightly higher compared to the step height of "as-delivered" surfaces (Figure \ref{fig:BufferLayerFormationOnSmallMiscut}(a)) and a noticeable improvement compared to the \SI{0.75}{nm} steps that are typically obtained by the standard etching procedure at \SI{1400}{\degreeCelsius} (Figure \ref{fig:BufferLayerFormationOnHydrogenEtchedSiC}(a)). After etching, the samples were applied to post-annealing at \SI{1175}{\degreeCelsius} for at least \SI{30}{min} to desorb adsorbed hydrogen from the SiC substrate which is accompanied by a further restructuring of the surface as shown in Figure \ref{fig:PASGGrapheneOnHydrogenEtchedSiC}(a). Additional information about hydrogen etching and post-annealing is given in the supplementary information. Polymer-assisted sublimation growth (PASG) was applied to support uniform buffer layer nucleation \cite{Kruskopf2016}. The deposition of polymer adsorbates was realized by liquid phase deposition (LPD) of AZ5214E photoresist (see ref. \cite{Kruskopf2016}). To control the size distribution of the adsorbate with heights $\leq \SI{2}{nm}$ shown in Figure \ref{fig:BufferLayerFormationOnPASG_H_etched_SiC}(a) the sample was purged in an ultrasonic bath of isopropanol. The higher adsorbate density used for the experiment given in Figure \ref{fig:BufferLayerFormationOnPASG_H_etched_SiC}(c) was realized by applying the procedure for the moderate density in the first step as well as by spin-coating of a weak solution of AZ5214E photoresist (\SI{6000}{rpm}, 4 droplets from the pipet solved in \SI{50}{ml} isopropanol) in the second step. The samples were introduced into the inductively heated hot-wall reactor which is typically evacuated to a base pressure of $\leq \SI{e-6}{mbar}$ before starting the process. The buffer layer samples were processed by annealing at \SI{1400}{\degreeCelsius} in argon at atmospheric pressure. To initiate graphene growth the temperature was further increased to \SI{1750}{\degreeCelsius} for 6 minutes.

\section*{Giant step bunching induced by the buffer layer reconstruction}

\begin{figure}
		\centering
    \includegraphics[width=1\textwidth]{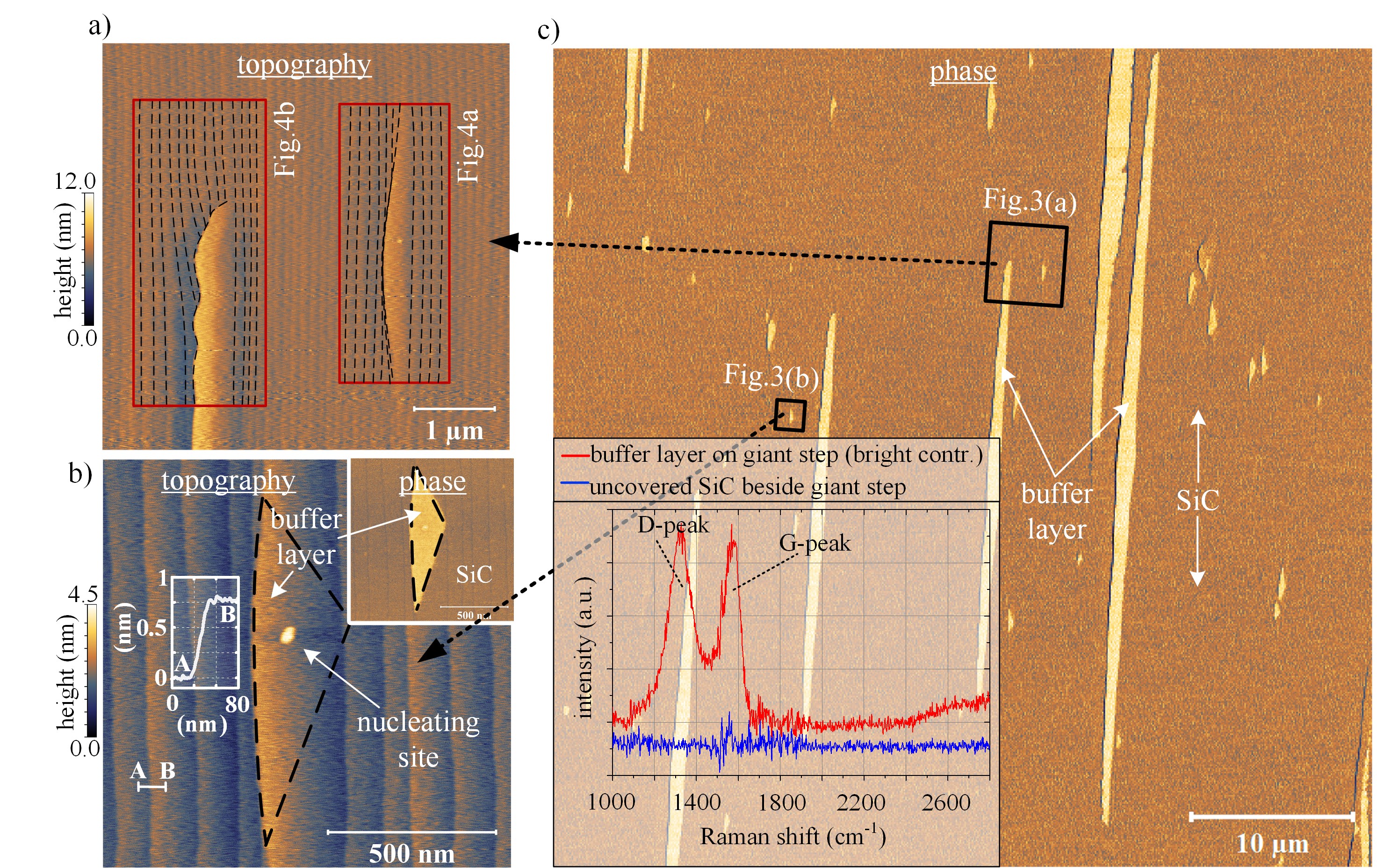}
	  \caption[Buffer layer and giant step formation on as-delivered large-miscut substrate]{\textbf{Buffer layer and giant step formation on the as-delivered large-miscut substrate.} \textbf{(a)} The AFM topography shows the initial stage of giant step bunching (right red box) as well as a further developed stage (left red box) in the vicinity of narrow and low terraces. The dotted black lines in the marked area indicate terrace edges. \textbf{(b)} Nucleation of giant steps is typically traced back to sites where the buffer layer growth occurs locally. The triangularly shaped buffer layer domain is identified by the light contrast in the corresponding phase image. The profile shows that narrow steps in the vicinity of giant steps have a step height of \SI{0.75}{nm}. \textbf{(c)} The large AFM phase image covers the areas shown in a) and b) as indicated. The Raman measurements in the inset show that buffer layer only forms on giant steps (corresponding to the light contrast) while the stable narrow steps (dark contrast) are uncovered.}
		\label{fig:BufferLayerFormationOnLargeMiscutSiC}
	\end{figure}

In the first experiment; the beginning of giant step formation on the SiC surface during high-temperature annealing (\SI{1400}{\degreeCelsius}, \SI{15}{ min}) in 1 bar argon atmosphere is studied on as-delivered substrates with a relatively large miscut angle as sketched in Figure \ref{fig:StartingSurfaces}(a). The atomic force microscopy (AFM) images in Figure \ref{fig:BufferLayerFormationOnLargeMiscutSiC} show three different kinds of terraces. In the magnified topographic images in Figure \ref{fig:BufferLayerFormationOnLargeMiscutSiC}(a-b) one can identify narrow-stepped regions corresponding to terraces with a width of $\approx \SI{100}{nm}$ and a height of \SI{0.75}{nm} (see profile in Figure \ref{fig:BufferLayerFormationOnLargeMiscutSiC}(b)) surrounding giant steps of different shapes. While the configuration of the narrow steps corresponding to repeated bunches of three SiC crystal layers is stable during the initial phase of the restructuring, also giant step formation is observed at some sites. The larger terraces are up to \SI{20}{nm} high, a few micrometers wide and several tens of micrometers long. Less developed giant steps such as those shown in Figure \ref{fig:BufferLayerFormationOnLargeMiscutSiC}(b) are a few micrometers long and less than \SI{0.5}{\um} wide. Here, the step formation seems to be connected to a particle-like topographic signal in the center. This situation with terraces at different development stages is the beginning of completely giant stepped surfaces as it is observed after graphene growth, Figure \ref{fig:PASGGrapheneOnHydrogenEtchedSiC}(b). The AFM phase contrast in the inset of Figure \ref{fig:BufferLayerFormationOnLargeMiscutSiC}(b) and in Figure \ref{fig:BufferLayerFormationOnLargeMiscutSiC}(c) indicates a clear correlation between the light contrast and giant step formation while the darker contrast corresponds to regions with narrow steps. Raman measurements at such sites were performed with a spot size smaller \SI{1}{\um} to identify the reason for the distinctive signals. The two Raman spectra in the inset of Figure \ref{fig:BufferLayerFormationOnLargeMiscutSiC}(c) are difference spectra which were obtained by subtracting the spectrum of an unprocessed SiC reference sample. The measurements reveal the characteristic signal of the buffer layer on broad terraces (red spectrum) while narrow stepped terrace regions show no Raman signal (blue spectrum) other than that of clean SiC. This surface condition describes the initiation of giant step bunching at \SI{1400}{\degreeCelsius} and demonstrates that uncovered narrow steps become unstable once local buffer layer nucleation occurs.

\begin{figure}
		\centering	
    \includegraphics[width=1\textwidth]{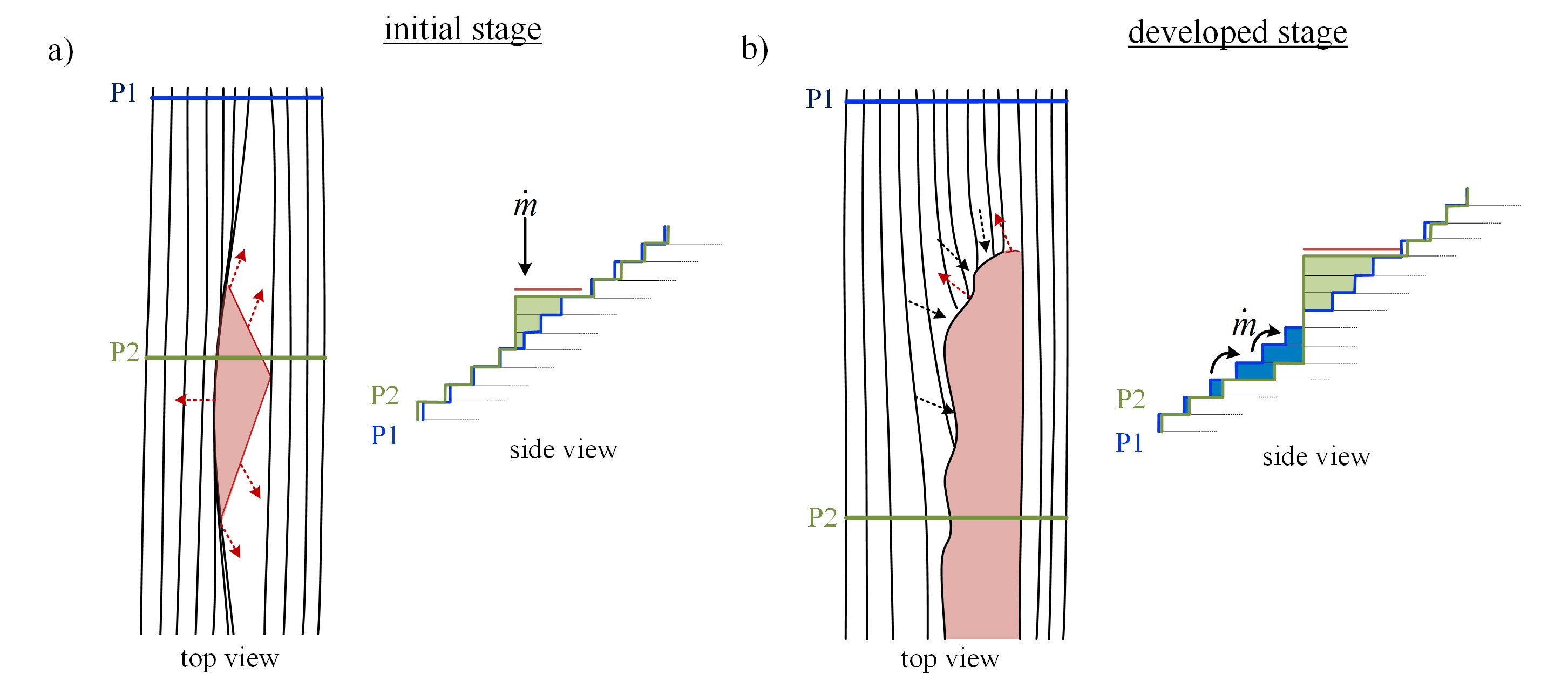}
	  \caption[Giant step bunching mechanisms]{\textbf{Giant step bunching mechanisms.} \textbf{(a)} The top view of the initial stage of giant step bunching depicts the step enlargement caused by the formation of a single buffer layer domain (red shade area with red arrows). The side view of the terrace structures created from the overlapping surface profile lines P1 (blue line) across the narrow steps and P2 (green line) across the giant step indicates the created volume (green shade) corresponding to the giant step. The uncorrelated step motion of the giant step with respect to the neighboring narrow steps suggests global mass transport $\dot{m}$ from distant terraces via the gas phase. \textbf{(b)} The top view of a developed giant step depicts the completely buffer layer covered terrace which represents an energetically favorable surface configuration. Such giant terraces continue to grow by material incorporation from neighboring narrow terraces as indicated by the black arrows. Here, the mass transport $\dot{m}$ is dominated by local transport which results in correlated terrace widths and movement of step edges. The overlapping profiles P1 and P2 on the right side indicate that the created volume of the giant step (green shade) is obtained by shifting material from neighboring uncovered terraces (blue shade).}
		\label{fig:GiantStepBunchingModel}
	\end{figure}

This behavior is explained by the general model developed by Jeong and Weeks. It states that the energetically favorable state of a new surface reconstruction is an important driving mechanism for step enlargement \cite{Weeks1997,Jeong1998,Jeong1995}. Our experimental results show that in the case of the SiC surface this reconstruction is represented by the buffer layer. Figure \ref{fig:GiantStepBunchingModel} depicts the top view of representative terraces and the corresponding side view of the step profiles created from the profile lines P1 (across narrow steps) and P2 (across the giant step). The drawings are derived from AFM measurements marked in Figure \ref{fig:BufferLayerFormationOnLargeMiscutSiC}(a). The depicted terrace regions covered by buffer layer (red shade area) visualize the correlation between the locally formed buffer layer and step enlargement. The process of giant step bunching can be divided into two stages: the initial phase of step nucleation (Figure \ref{fig:GiantStepBunchingModel}(a)) and the developed phase of continued growth of giant steps (Figure \ref{fig:GiantStepBunchingModel}(b)). Despite the long annealing time of \SI{15}{min}, many of the evolving broader terraces did not reach the stage of fully developed giant steps which suggests that the mechanisms in the initial growth stage are relatively slow. Regarding the model of Jeong and Weeks, this is typically the case as long as the width of a terrace is insufficiently wide to create a so-called “critical nucleus” \cite{Weeks1997,Jeong1998,Jeong1995}. On the SiC surface, one needs to distinguish between (i) a growing buffer layer domain that slowly enforce a widening of the terrace in the first stage (Figure \ref{fig:GiantStepBunchingModel}(a)) and (ii) in the second stage the relatively fast growing giant steps when the critical width is reached (Figure \ref{fig:GiantStepBunchingModel}(b)). Both processes (buffer layer and giant step formation) have their individual critical nucleus size. Critical nuclei for forming local buffer layer domains (leading to the stage (i)) are expected to be sometimes related to surface contamination e.g. remaining particles after cleaning (derived from Figure \ref{fig:BufferLayerFormationOnLargeMiscutSiC}(b)). A giant step of critical size as depicted in Figure \ref{fig:GiantStepBunchingModel}(b) has an estimated step height of at least 3 unit cell heights (\SI{4.5}{nm}) which corresponds to a width of $\geq \SI{0.6}{\um}$ in the case of the large-miscut substrate (derived from Figure \ref{fig:BufferLayerFormationOnLargeMiscutSiC}).

The growth of the buffer layer implies carbon supersaturation at the SiC surface. This is understandable since the much higher vapor pressure of silicon supports enhanced silicon desorption while the low vapor pressure of carbon species implies a relatively low desorption rate leading to carbon enrichment. Thus, the surface conditions during buffer layer growth may be described by nucleation in the presence of a “sea” of diffusing carbon adatoms. The growth of the buffer layer islands causing giant step bunching seems to suppress the formation of others in their vicinity within a distance of several micrometers (approximately \SI{3}{\um} to \SI{4}{\um} in Figure \ref{fig:BufferLayerFormationOnLargeMiscutSiC}). In the literature this behavior is described as being typical when a single nucleus locally reduces the level of supersaturation, thus preventing the formation of other stable domains in the nearby region \cite{Dhanaraj2010,Dubrovskii2014,Lui1999}. Considering a mean terrace width of about \SI{0.1}{\um}, this implies transport of carbon species across several step edges. Considering a mean terrace width of about \SI{0.1}{\um} this implies transport of carbon species across several edges.

The nucleation behavior and the uncorrelated and correlated motion of steps during giant step formation suggest the presence of mass transport processes that match the description of the general model presented in Figure 1. Depending on the development stage of SiC terraces the characteristics of both local and global transport mechanisms were identified for the samples annealed in an argon atmosphere. Evidence of global transport via the crystal/vapor interface is given by the step/terrace structure during the initial phase of giant step formation as depicted in Figure \ref{fig:GiantStepBunchingModel}(a). Here, the growth of a surface-reconstruction-induced giant step is uncorrelated with the shape of neighboring terraces due to mass transport from distant terraces via desorption and adsorption. The surface profiles P1 (blue profile) across the narrow steps and P2 (green line) across the giant step on the right side of Figure \ref{fig:GiantStepBunchingModel}(a) indicate that a substantial amount of material is transported towards the nucleation site via the vapor phase denoted by $\dot{m}$. The material to create the giant step is indicated by the green shade area which is a result of the overlapping of P1 and P2. A different situation was identified in the case of developed giant steps as depicted in Figure \ref{fig:GiantStepBunchingModel}(b). Here, the shape and width of uncovered neighboring terraces correlate with the growth of the giant step which is a characteristic of local mass transport processes. In this case, the mass transport $\dot{m}$ mainly occurs due to surface diffusion between neighboring step edges. Due to different retraction velocities, the narrow terraces close to the giant step become broader than more distant ones. This situation can also be understood from the corresponding profile lines P1 (blue line) across the narrow steps and P2 (green line) across the giant step shown on the right side of Figure \ref{fig:GiantStepBunchingModel}(b). In this case, the material to create the volume of the giant step (green shade) is obtained via surface diffusion from neighboring uncovered terraces (blue shade). Note that the blue shaded area was also for other profile positions (not shown) slightly smaller than the green area which also implies a contribution of global transport processes in the case of developed steps. Further details about giant step formation on small-miscut substrates are given in Figure \ref{fig:BufferLayerFormationOnHydrogenEtchedSiC}(b).

\section*{Step-nucleation on small-miscut substrate - a comparison to larger miscut angles}

\begin{figure}
		\centering
    \includegraphics[width=1\textwidth]{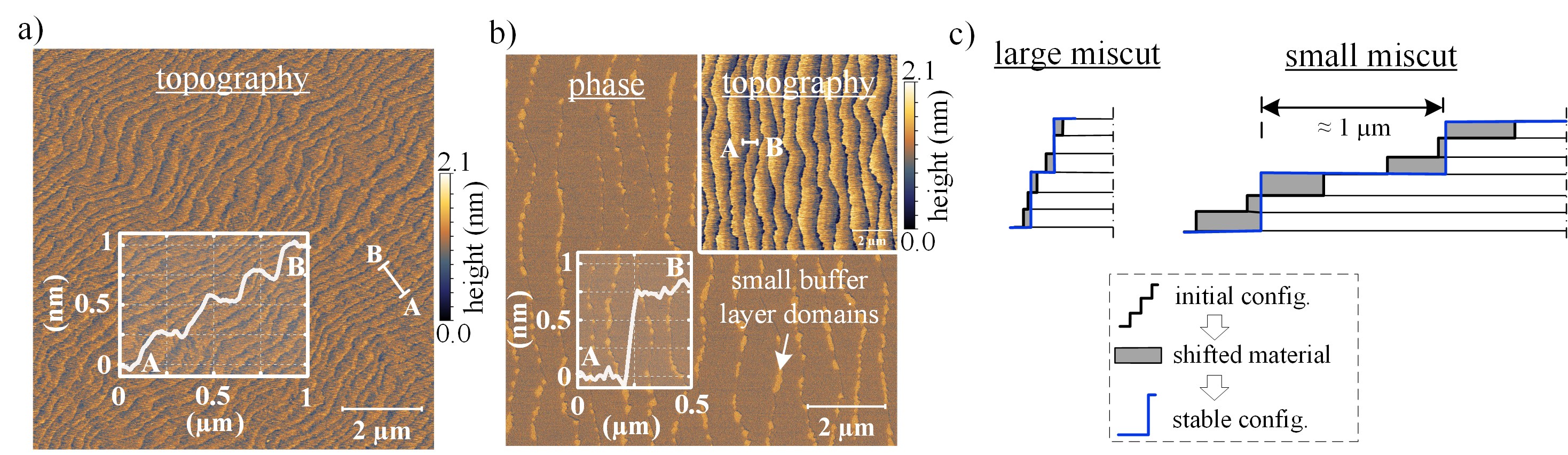}
	  \caption[Buffer layer step-nucleation on small-miscut substrate]{\textbf{Buffer layer growth by step-nucleation on "as-delivered" small-miscut substrate.} \textbf{(a)} AFM topography image of a 6H-SiC substrate with a miscut of $\approx \SI{0.05}{\degree}$. in the as-delivered state from the manufacturer. The CMP (chemical mechanical polished) starting surface has uniform step and terrace structures with heights of \SI{0.25}{nm} (see profile) corresponding to single SiC crystal layers. \textbf{(b)} Restructuring of the clean surface results in a stable surface configuration with step heights of \SI{0.75}{nm} (see profile) corresponding to three SiC bilayers after argon-annealing at \SI{1400}{\degreeCelsius}. The AFM phase image identifies uniform buffer layer islands that aligned along the lower side of step edges. \textbf{(c)} The simplified representations of the surface profiles of large-miscut (left) and small-miscut substrate (right) sketch the conversion of the initial step configuration with single SiC crystal layers (black outline) into the stable step configuration (blue outline). This conversion does not necessarily involve local buffer layer formation.}
		\label{fig:BufferLayerFormationOnSmallMiscut}
	\end{figure}

One approach to slow the decomposition of the SiC surface and to prevent giant step bunching is to predefine continuous buffer layer domains using substrates with very low-miscut angles \cite{Virojanadara2009,Virojanadara2010,Kruskopf2015}. In this experiment, the annealing procedure at \SI{1400}{\degreeCelsius} in an argon atmosphere was applied using as-delivered SiC substrate with a small-miscut angle. The AFM image in Figure \ref{fig:BufferLayerFormationOnSmallMiscut}(a) shows the starting surface configuration with single SiC bilayer steps. After annealing regular steps with heights of \SI{0.75}{nm} evolve, see Figure \ref{fig:BufferLayerFormationOnSmallMiscut}(b). The phase image reveals that in addition to SiC (dark contrast), small buffer layer domains (light contrast) form along the energetically preferred lower side of the edges. Steps lower than \SI{0.75}{nm} were not observed since they are not stable and decompose quickly during the initial phase of surface restructuring.

The results imply that during annealing at \SI{1400}{\degreeCelsius} in argon atmosphere silicon and carbon species may exchange between terraces enabling the formation of a new step configuration and only partly contributing to the buffer layer growth. While this holds also true for the large-miscut substrate the nucleation of the buffer layer is significantly different. This demonstrates that the miscut angle is a critical parameter that strongly influences the dynamics of the restructuring processes. 
The simplified schematics of surface profiles of large (left) and small (right) miscut surfaces in Figure \ref{fig:BufferLayerFormationOnSmallMiscut}(c) describe the thermally activated conversion of the starting surface configuration (black outline, \SI{0.25}{nm} steps) into the energetically preferred stable configuration (blue outline, \SI{0.75}{nm} steps). The comparison shows that as a result of broader terraces on a small-miscut surface a bigger net mass transfer (gray shaded area) is involved in fulfilling the conversion. The higher amount of released carbon is expected to be the reason for the increased buffer layer coverage and the reduced tendency to form giant steps.

\section*{Giant step bunching and buffer layer growth on hydrogen-etched substrate}

\begin{figure}
		\centering
    \includegraphics[width=1\textwidth]{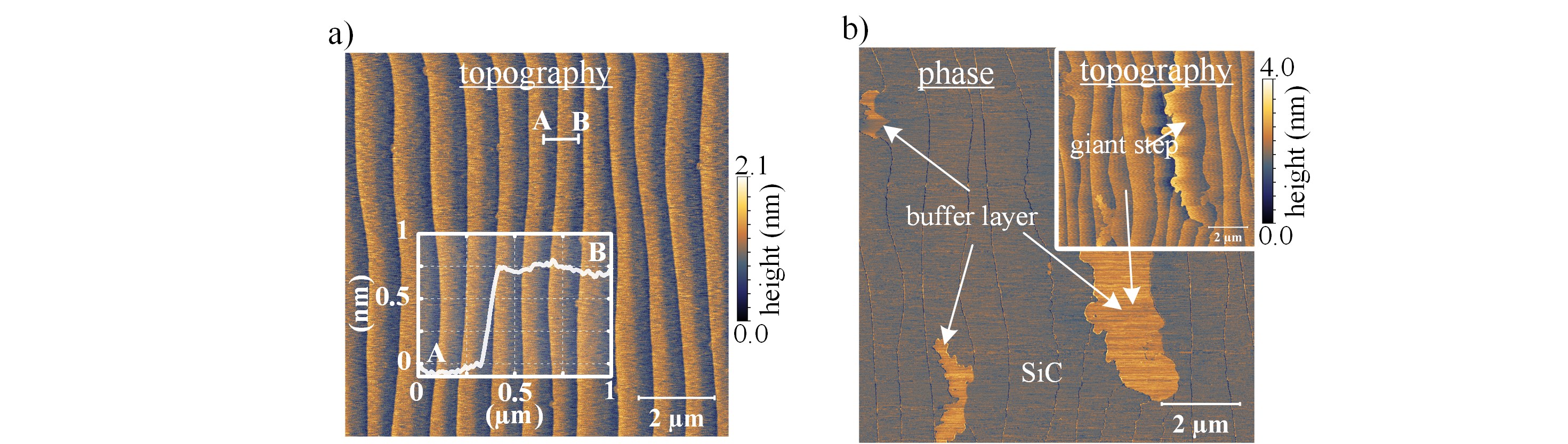}
	  \caption[Buffer layer growth on the hydrogen-etched small-miscut substrate]{\textbf{Buffer layer formation on hydrogen-etched small-miscut substrate.} \textbf{(a)} AFM topography image of the starting SiC surface shows regular terraces, and \SI{0.75}{nm} steps (see profile) that were realized by hydrogen etching at \SI{1400}{\degreeCelsius}. \textbf{(b)} After annealing at \SI{1400}{\degreeCelsius} the AFM image shows the formation of a few buffer layer domains (light contrast) leading to giant step formation (see topography in the inset).}
		\label{fig:BufferLayerFormationOnHydrogenEtchedSiC}
\end{figure}

In the following set of experiments, the initial phase of surface restructuring is skipped by using hydrogen-etched samples that were processed at the standard etching temperature of \SI{1400}{\degreeCelsius} as sketched in Figure \ref{fig:StartingSurfaces}(c). The corresponding topography in Figure \ref{fig:BufferLayerFormationOnHydrogenEtchedSiC}(a) shows that in this way the starting surface of the "as-delivered" epi-ready wafer (see Figure \ref{fig:BufferLayerFormationOnSmallMiscut}(a)) is converted into regular bunches of 3 SiC bilayers before the buffer layer growth is initiated. By skipping the restructuring process, its influence on the buffer layer formation can be analyzed. First, a low-temperature post-annealing process step is necessary to desorb remaining hydrogen from the substrate after the hydrogen etching process. Without this process, any subsequent process involving buffer layer or graphene growth is significantly delayed or can be very nonuniform, see supplementary data. The etching and post-annealing procedure results in an extremely clean, (1x1) reconstructed SiC surface with uniform and {0.75}{nm} high steps in agreement with AFM (Figure \ref{fig:BufferLayerFormationOnHydrogenEtchedSiC}(a)) and low energy electron diffraction (LEED) (data not shown).

Figure \ref{fig:BufferLayerFormationOnHydrogenEtchedSiC}(b) shows the buffer layer growth on the hydrogen-etched substrate. While most of the terraces remain uncovered and preserve their heights, nucleation of the buffer layer occurs locally at a few randomly distributed sites that are separated by several micrometers. The formation of these sites is accompanied by local step enlargement and further giant step bunching once the step height reaches about 1.5 unit cell heights (2.25 nm). This corresponds to an estimated critical width of about \SI{2}{\um} for the small-miscut substrate which is nearly a factor of four wider compared to the critical width identified on large-miscut substrates. From the comparison with the as-delivered small-miscut substrate without applying hydrogen etching (Figure \ref{fig:BufferLayerFormationOnHydrogenEtchedSiC}) one can conclude that skipping the restructuring process of the SiC surface leads to non-uniform buffer layer nucleation during annealing at \SI{1400}{\degreeCelsius} in an argon atmosphere. One possible reason for this effect is the absence of nuclei due to the extreme cleanness of the etched surface. Additionally, less carbon is released since the step configuration is already stable and no initial restructuring takes place. Therefore, the first domains must have formed by spontaneous nucleation once a critical level of carbon supersaturation is reached. As it was already found in the case of the large-miscut substrate (Figure \ref{fig:BufferLayerFormationOnLargeMiscutSiC}), the formation of each stable domain seems to suppress the formation of others in their vicinity by locally reducing the level of supersaturation.

\section*{Buffer layer nucleation and suppression of giant step formation by surface polymer treatment}

\begin{figure}
		\centering
    \includegraphics[width=1\textwidth]{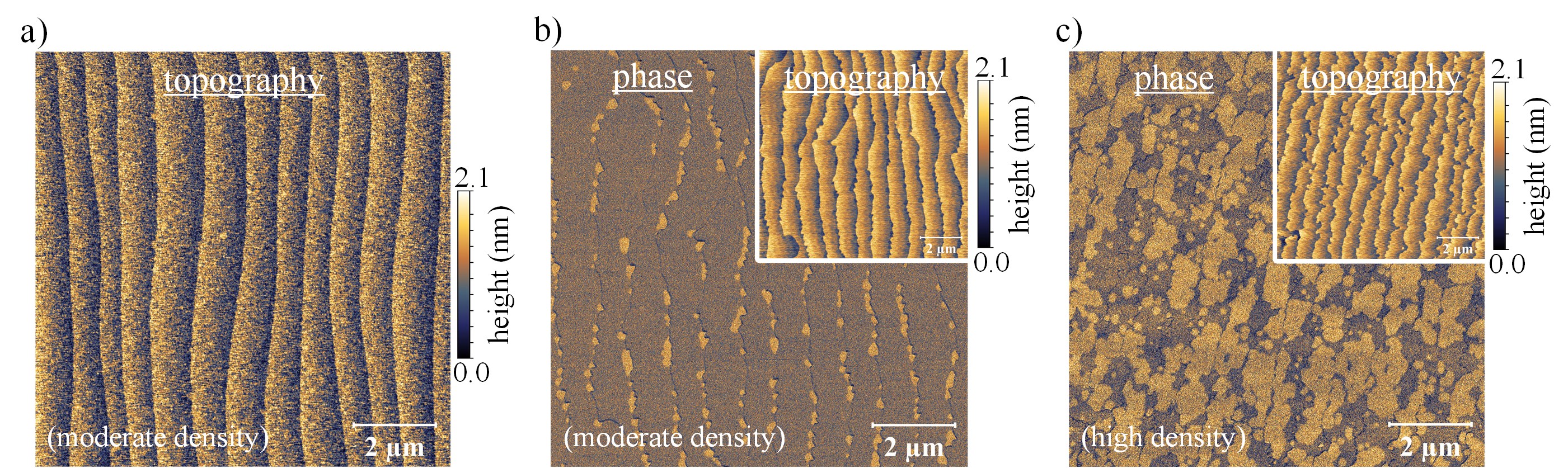}
	  \caption[Polymer-assisted buffer layer growth on hydrogen-etched substrates.]{\textbf{Polymer-assisted buffer layer growth on hydrogen-etched substrates.} \textbf{(a)} AFM topography image of the hydrogen-etched and polymer-treated starting surface with steps of \SI{0.75}{nm} shows the deposited adsorbate with structure heights of $\approx \SI{2}{nm}$ before annealing.  \textbf{(b)} Argon-annealing of the PASG treated surface at \SI{1400}{\degreeCelsius} leads to enhanced nucleation of buffer layer domains compared to untreated surfaces and helps to preserve the initial step height. The nucleation behavior in the case of a moderate density of polymer follows the description of step-nucleation. \textbf{(c)} Deposition of a higher density of the polymer adsorbate increases the coverage and leads to terrace-nucleation.}
		\label{fig:BufferLayerFormationOnPASG_H_etched_SiC}
\end{figure}

The following set of experiments is a direct comparison to the results shown in Figure \ref{fig:BufferLayerFormationOnHydrogenEtchedSiC} using the same annealing procedure as well as hydrogen-etched SiC surfaces. The samples presented in Figure \ref{fig:BufferLayerFormationOnPASG_H_etched_SiC} were additionally treated with a polymer adsorbate which supports the buffer layer formation. The principle of this so-called polymer-assisted sublimation growth (PASG) method \cite{Kruskopf2016} is to control the amount of available carbon and related nuclei as described in the sample preparation. The starting surface of the first sample shown in Figure \ref{fig:BufferLayerFormationOnPASG_H_etched_SiC}(a) was modified with a moderate adsorbate density resulting in increased structure heights of $\approx \SI{2}{nm}$ (Figure \ref{fig:BufferLayerFormationOnPASG_H_etched_SiC}(a)).

Indeed, after annealing at \SI{1400}{\degreeCelsius} in an argon atmosphere, the topography of the two PASG buffer layer samples (Figure \ref{fig:BufferLayerFormationOnPASG_H_etched_SiC}(b-c)) showed no giant step bunching which is in contrast to the hydrogen-etched substrate without polymer treatment (Figure \ref{fig:BufferLayerFormationOnHydrogenEtchedSiC}(b)). However, the results are remarkably similar compared to that observed on the as-delivered small-miscut substrate shown in Figure \ref{fig:BufferLayerFormationOnSmallMiscut}(b) even though the starting surfaces were different before annealing and only the etched surface was treated with the polymer. On both surfaces uniformly distributed buffer layer domains formed along the lower side of each step edge and act as preferred nucleation sites. In a second experiment (Figure \ref{fig:BufferLayerFormationOnPASG_H_etched_SiC}(c)), a hydrogen-etched sample was treated with with a higher adsorbate density as described in the sample preparation. Due to the larger amount of available carbon, the buffer layer coverage significantly increased. 
These results show that carbon which is released during the restructuring process of the SiC substrate may be substituted by the carbon from the deposited polymer leading to enhanced nucleation and forming a high density of buffer layer domains. The seeded growth prevents the spontaneous formation of large and separated domains and thus the concomitant giant step bunching.

\section*{Step-edge and terrace-nucleation models}

\begin{figure}[t]
		\centering
    \includegraphics[width=1\textwidth]{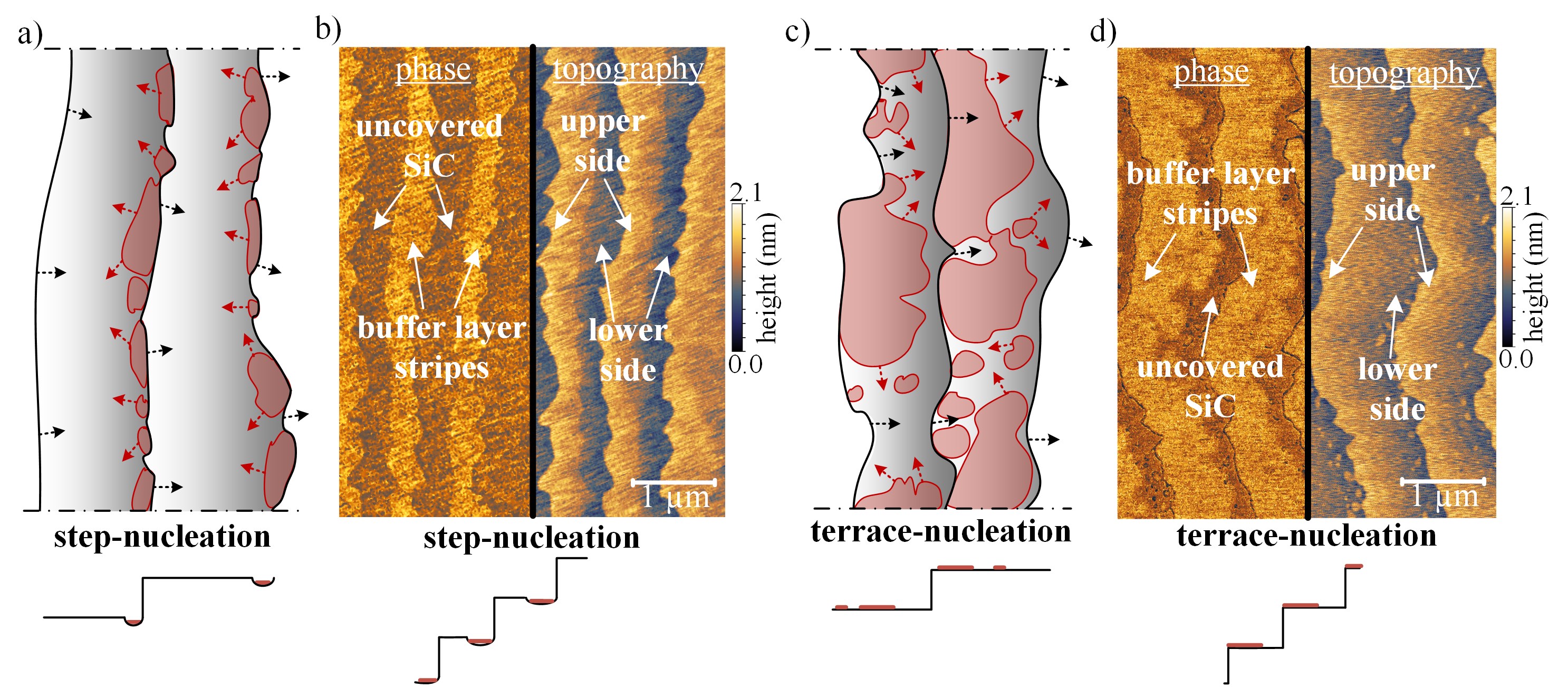}
	  \caption[Buffer layer nucleation mechanisms]{\textbf{Buffer layer nucleation mechanisms.} \textbf{(a-b)} In the case of moderate carbon supply typically “step-nucleation” of the buffer layer is observed. \textbf{(a)} The drawing is derived from Figure \ref{fig:BufferLayerFormationOnPASG_H_etched_SiC}(b) and describes the beginning of step-nucleation (red shade buffer layer) along the lower side of step edges. \textbf{(b)} The AFM phase and topography images show continuous domains that were realized by increasing the annealing time to 30 min fed by the decomposition of SiC step edges as well as of underlying crystal planes. \textbf{(c-d)} Extra carbon supply from the polymer adsorbate leads to terrace-nucleation. \textbf{(c)} The drawing is derived from Figure \ref{fig:BufferLayerFormationOnPASG_H_etched_SiC}(c) and describe the formation of buffer layer islands on the terraces. \textbf{(d)} The AFM phase and topography images show that a longer annealing time of 30 min leads to the coalescence of individual domains forming continuous stripes along the upper side of the terrace edges.}
		\label{fig:BufferlayerNucleationMechanisms}
\end{figure}

From the different shapes and the location of the buffer layer domains at terrace or edge sites one can distinguish between two different mechanisms of 2D-island growth (Figure \ref{fig:BufferLayerFormationOnSmallMiscut}(b) and Figure \ref{fig:BufferLayerFormationOnPASG_H_etched_SiC}(b-c)). Depending on the amount of available carbon a transition from one to the other type can be obtained.
The first is called “step-nucleation” (Figure \ref{fig:BufferlayerNucleationMechanisms}(a-b)) and describes the case where small domains form along the lower side of step edges. Based on the theory of heteroepitaxy, nucleation at step edges implies near equilibrium growth conditions with a diffusion length $\lambda_{dif}$ larger than the terrace width ($w_{terrace} \approx \SI{1}{\um}$) such that the species have sufficient time to migrate on the terrace until a stable edge position is found \cite{Hasselbrink2008,Lui1999}. The comparison of the morphology described in Figure \ref{fig:BufferlayerNucleationMechanisms}(a) and that of the AFM measurements in Figure \ref{fig:BufferlayerNucleationMechanisms}(b) shows increasing the annealing time by \SI{15}{min} leads to the formation of continuous buffer layer stripes (light phase contrast) along the lower side of the edges from where they continue to grow onto the terrace. The measurements indicate that the growing domains are fed from decomposing step edges as well as from crystal layers from underneath since they are typically located slightly lower than the height level of the SiC surface as indicated in the step profiles. 
The second type of 2D-layer growth shown in Figure \ref{fig:BufferlayerNucleationMechanisms}(c-d) is called “terrace-nucleation”  and describes the case where buffer layer domains form on the terraces and not along the step edges. Based on theory, nucleation on the terrace is obtained under supersaturated conditions once the effective diffusion length $\lambda_{dif}$ becomes smaller than the terrace width \cite{Hasselbrink2008,Lui1999}. On the SiC surface, the reduction of $\lambda_{dif}$ leading to a transition from step-nucleation to terrace-nucleation of buffer layer nuclei is achieved by increasing the amount of the deposited carbon. Figure \ref{fig:BufferlayerNucleationMechanisms}(c) depicts the morphology identified in Figure \ref{fig:BufferLayerFormationOnPASG_H_etched_SiC}(c) with randomly distributed domains. The sketched surface profiles (below) and the AFM images in Figure \ref{fig:BufferlayerNucleationMechanisms}(d) show that in the case of longer annealing times the distributed islands coalesce and form continuous domains which align along the upper side of the terrace edges.

\section*{Graphene formation on hydrogen-etched polymer-treated surfaces}

\begin{figure}
  \centering
  \includegraphics[width=1\textwidth]{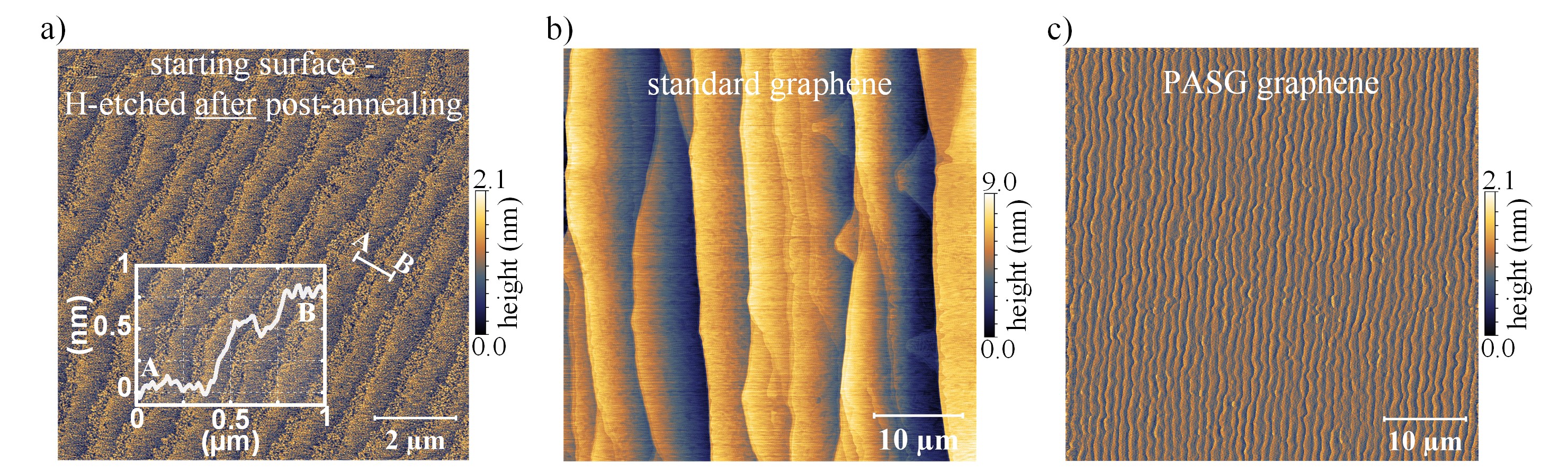}
	\caption[PASG graphene on hydrogen-etched substrates]{\textbf{AFM comparison of graphene growth on the hydrogen-etched substrates with and without polymer treatment.} \textbf{(a)} The hydrogen-etched surface shows regular terraces with step heights of \SI{0.25}{nm} and \SI{0.5}{nm} (see profile) after etching at \SI{1200}{\degreeCelsius} and post-annealing at \SI{1175}{\degreeCelsius}. \textbf{(b)} During standard graphene growth without PASG treatment giant steps develop with structure height of up to $\approx \SI{10}{nm}$. \textbf{(c)} Graphene growth by the PASG method results in uniform step and terrace structures with heights $\leq \SI{0.75}{nm}$. Both polymer-treated and untreated substrates were processed at the same time.}
\label{fig:PASGGrapheneOnHydrogenEtchedSiC}
\end{figure}

The significance of the predefined buffer layer on the graphene growth process can be well demonstrated using simultaneously processed hydrogen-etched surfaces. Compared to the etched surface in Figure \ref{fig:BufferLayerFormationOnHydrogenEtchedSiC}(a) the step height of \SI{0.75}{nm} was reduced by applying a lower etching temperature of \SI{1200}{\degreeCelsius} such that a sequence of two steps with heights of \SI{0.25}{nm} and \SI{0.5}{nm} are obtained as shown in Figure \ref{fig:PASGGrapheneOnHydrogenEtchedSiC}(a). This incompletely restructured surface is expected to support the buffer layer growth in agreement with the previous experiments.
However, graphene growth without polymer treatment (Figure \ref{fig:PASGGrapheneOnHydrogenEtchedSiC}(b)) led to broad terrace structures with heights of up to \SI{10}{nm}. Compared to the result after buffer layer growth (Figure \ref{fig:BufferLayerFormationOnHydrogenEtchedSiC}(b)), giant step bunching is significantly enhanced during graphene growth at $T = \SI{1750}{\degreeCelsius}$. Thus, the slightly reduced step heights and the buffer layer formed during annealing in argon atmosphere alone cannot sufficiently stabilize the surface. This behavior is typical for hydrogen-etched substrates \cite{Emtsev2009,Oliveira2011}. The previous experiments in Figure \ref{fig:BufferLayerFormationOnPASG_H_etched_SiC}(b-c) showed that the application of polymer adsorbates successfully circumvent giant step bunching at \SI{1400}{\degreeCelsius}. The PASG graphene sample shown in Figure \ref{fig:PASGGrapheneOnHydrogenEtchedSiC}(c) proves that this also holds true for high temperatures. The slightly reduced initial step heights of the etched surface of \SI{0.25}{nm} and \SI{0.5}{nm} lead to small but noticeable improvements in the final topography of the graphene sample e.g. higher terrace uniformity and reduced step heights compared to a hydrogen-etched starting surface with exclusively \SI{0.75}{nm} (not shown). This is expected to be a consequence of the incomplete restructuring process of the surface before the buffer layer, and graphene growth processes are initiated. The realization of large-area growth of monolayer graphene with step heights $\leq \SI{0.75}{nm}$ on hydrogen-etched substrates demonstrates the versatility of the PASG method and underlines the high importance of the buffer layer formation.

\section*{Conclusion}

Understanding the morphological changes that occur during the restructuring of the SiC surface lead to significant improvements in the concept of large-area graphene growth. The identified critical point is the conversion of the starting surface into the next stable configuration which is usually accompanied by 2D-island growth of the buffer layer. Three different growth mechanisms were identified namely (i) surface-reconstruction-induced giant step formation as well as buffer layer formation by (ii) step-nucleation or (iii) terrace-nucleation. Their occurrence is closely connected to the properties of the surface and the selected process conditions e.g. the carbon supply. Surface-reconstruction-induced giant step bunching usually occurs when the initial restructuring process of the starting surface involves only a small amount of carbon as it was demonstrated using clean as-delivered large-miscut or hydrogen-etched substrates. Here, the high mobility mass transport between adjacent terraces involving surface diffusion as well as between distant terraces involving desorption and adsorption favors the formation of separated buffer layer domains. The observed mechanisms reveal that the widening of covered terraces is due to the locally created energetically favorable surface configuration of the buffer layer reconstruction. This behavior can be prevented throughout the whole graphene formation process if a high density of small buffer layer domains is grown using suitable substrates and an additional carbon source. Moderate carbon supply favors step-nucleation along the lower side of step edges and usually results in a low coverage. Increasing the amount of available carbon, however, realizes a transition from step-nucleation to terrace-nucleation. Terrace-nucleation under highly supersaturated conditions was identified to be the ideal growth mode since it enables to conserve the lowest possible step heights throughout the whole graphene process and extends the range of suitable substrates. The manipulation of the growth dynamics by seeded buffer layer growth suppresses critical step bunching mechanisms and makes uniform buffer layer growth also possible for hydrogen etched substrates. These results are of particular importance for the large-area growth of monolayer graphene on SiC with ultra-low steps.

\section*{Acknowledgements}
We gratefully acknowledge funding by the \textit{School for Contacts in Nanosystems (NTH nano)} and the support by the \textit{Braunschweig International Graduate School of Metrology (B-IGSM)} and \textit{NanoMet}.

\section*{References}
\label{refs}
\bibliography{StudyOnTheMorphologyAndGrowth_Main_Supplementary}



\section*{Supplementary material}

# Supplementary information:

## Tailoring the SiC surface - a morphology study on the epitaxial growth of graphene and its buffer layer

**Mattias Kruskopf[1](*), Klaus Pierz[1], Davood Momeni Pakdehi[1], Stefan Wundrack[1], Rainer Stosch[1], Andrey Bakin[2,3] and Hans W. Schumacher[1]**

[1]Physikalisch-Technische Bundesanstalt, Bundesallee 100, 38116 Braunschweig, Germany

[2]Institute of Semiconductor Technology of Technische Universität Braunschweig, Hans-Sommer-Straße 66, 38106 Braunschweig, Germany

[3]Laboratory for Emerging Nanometrology (LENA), TU Braunschweig, Germany

E-mail: `Mattias.Kruskopf@ptb.de, Klaus.Pierz@ptb.de`

June 2017

## Low-temperature post-annealing of H-etched substrates

Hydrogen etching of SiC is a widely used preparation technique to clean the surface from adsorbates e.g. oxides and to remove polishing damages by predefining regularly stepped surface facets at temperatures between $1200\,°C$ to $1700\,°C$ [1–5]. Etching at relatively low temperatures of $1200\,°C$ leads to step heights of $0.25\,nm$ and $0.5\,nm$ as shown in Figure S1(a). This is only slightly higher compared to the step height of an "as-delivered" surface given in Figure 5(a) and thus is an improvement to the standard etching procedure at $1400\,°C$ which results in higher steps of $0.75\,nm$ shown in Figure 6(a). However, not only the topography changes during the etching process. It was observed that the etched substrate behaved very differently compared to as-delivered surfaces especially during polymer-assisted growth. Figure S1(b) demonstrates the growth result using the standard PASG graphene process on hydrogen etched substrate that was not introduced to post-annealing which is in strong contrast to the results shown in Figure 9(c). Without post-annealing, the layer formation is very non-uniform. It was identified that post-annealing of the substrate is mandatory before the polymer treatment. Note, that the post-annealing procedure applied in this work ($1175\,°C$, $1\,bar$ Ar atmosphere) leads to a further restructuring of the surface (see Figure 9(a)). These experimental results support the understanding that adsorbed hydrogen needs to be desorbed from the SiC surface before the growth process is initiated. Hydrogen desorption and adsorption on the SiC surface and diffusion inside the bulk at temperatures $T \geq 1100\,°C$ is well described in the literature [6, 7]. However, adsorbed hydrogen is not reported as a problem

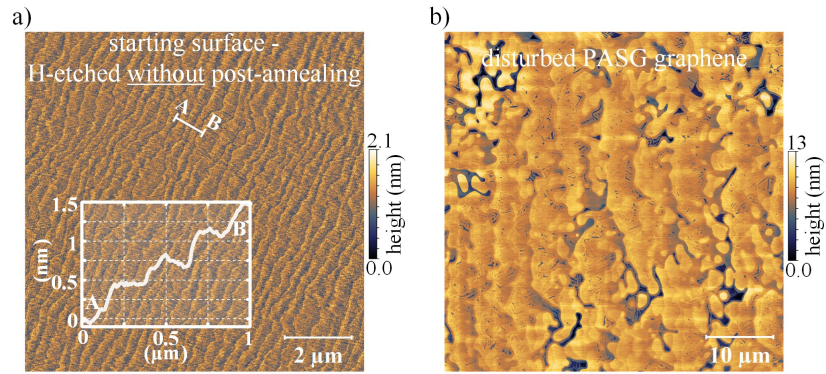

**Figure S1. The importance of low-temperature post-annealing using H-etched substrates. (a)** SiC substrate after hydrogen etching at 1200 °C without post-annealing shows regular steps with heights of 0.25 nm and 0.5 nm. **(b)** Graphene growth on hydrogen etched substrate without post-annealing may lead to unpredictable growth results. This is probably due to adsorbed hydrogen at the substrate surface which needs to be desorbed to not disturb the buffer layer and graphene growth.

in the literature about standard step-flow grown graphene [8–10]. This is understandable since these processes usually apply no separate buffer layer growth step but directly initiate the graphene growth at higher temperature ($\geq$ 1650 °C) where the hydrogen desorbs much faster. The developed processes in this work though, predefine a buffer layer on shallow SiC terraces at relatively low temperatures ($\approx$ 1400 °C) supported by the additional carbon source from the polymer. At this growth stage, adsorbed hydrogen is expected to react with the carbon atoms on the surface and to disturb the growth. Understanding that post-annealing after hydrogen etching is a prerequisite for controlled buffer layer growth is an important contribution to the PASG method.



\end{document}